# A SURVEY of RECENT INTRUSION DETECTION SYSTEMS for WIRELESS SENSOR NETWORK

Tapolina Bhattasali [1], Rituparna Chaki [2]

[1]Techno India College of Technology, Kolkata, India

tapolinab@gmail.com

[2]West Bengal University of Technology, Kolkata, India

rituchaki@gmail.com

**Abstract.** Security of Wireless sensor network (WSN) becomes a very important issue with the rapid development of WSN that is vulnerable to a wide range of attacks due to deployment in the hostile environment and having limited resources. Intrusion detection system is one of the major and efficient defensive methods against attacks in WSN. A particularly devastating attack is the sleep deprivation attack, where a malicious node forces legitimate nodes to waste their energy by resisting the sensor nodes from going into low power sleep mode. The goal of this attack is to maximize the power consumption of the target node, thereby decreasing its battery life. Existing works on sleep deprivation attack have mainly focused on mitigation using MAC based protocols, such as S-MAC, T-MAC, B-MAC, etc. In this article, a brief review of some of the recent intrusion detection systems in wireless sensor network environment is presented. Finally, we propose a framework of cluster based layered countermeasure that can efficiently mitigate sleep deprivation attack in WSN. Simulation results on MATLAB exhibit the effectiveness of the proposed model in detecting sleep-deprivation attacks.

## 1 Introduction

Wireless sensor network (WSN) refers to a system that consists of number of low-cost, resource limited sensor nodes to sense important data related to environment and to transmit it to sink node that provides gateway functionality to another network, or an access point for human interface. WSN is a rapidly growing area as new technologies are emerging, new applications are being developed, such as traffic, environment monitoring, healthcare, military applications, home automation. WSN is vulnerable to various attacks such as jamming, battery drainage, routing cycle, sybil, cloning. Due to limitation of computation, memory and power resource of sensor nodes, complex security mechanism can not be implemented in WSN. Therefore energy-efficient security implementation is an important requirement for WSN.

A sleep deprivation attack (battery drainage) is a particularly severe attack in WSN because recharging or replacing node batteries in WSN may be impossible. In this type of attack, intruder forces the sensor nodes to remain awake; so that they waste their energy. This attack imposes such a large amount of energy consumption upon the limited power sensor nodes that they stop working and give rise to denial of service through denial of sleep.

In this paper a survey of on-going research activity is presented. This is followed by a comparative analysis of the recent ID schemes. This paper concludes with a glimpse of the proposed model for detecting sleep deprivation attack.

## 2   Related Works

Intrusion detection for WSN is an emerging field of research. This section presents a category-wise report of on-going research activities.

**Distributed Approach**

- In [1], a semantic based intrusion detection framework is proposed for WSN by using multi-agent and semantic based techniques, where security ontology is constructed according to the features of WSN to represent the formal semantics for intrusion detection. This distributed technique is based on cooperative mechanism. In this mechanism, each selected rule of security ontology is mapped to sensing data collected from common sensor nodes to detect anomaly.
- In [2], an energy efficient learning solution for IDS in WSN has been proposed. This schema is based on the concept of stochastic learning automata on packet sampling mechanism. Simple Learning Automata based ID (S-LAID) functions in a distributed manner with each node functioning independently without any knowledge about the adjacent nodes.

**Hierarchical Approach**

- In [3], a location-aware, trust-based detection and isolation mechanism of compromised nodes in wireless sensor network is proposed. In this technique, probabilistic model is used to define trust and reputation.
- In [4], a method using isolation table is proposed to isolate malicious nodes by avoiding consumption of unnecessary energy by IDS (ITIDS).This hierarchical structure of IDS based on cluster network can detect serious attacks such as hello flooding, denial of service (DoS), denial of sleep, sinkhole and wormhole attack. In this mechanism, malicious nodes can be detected by considering remaining energy and trust values of sensor nodes.
- In [5], a lightweight ranger based IDS (RIDS) has been proposed. It combines the ranger method to reduce energy consumption and the isolation tables to avoid detecting anomaly repeatedly. This lightweight IDS model relates ontology concept mechanism about anomaly detection. In this technique, rough set theory (RST) is used for preprocessing of packets and anomaly models will be trained by support vector machine (SVM).
- In [6], a hierarchical overlay design (HOD) based intrusion detection system is proposed, using policy based detection mechanism. This model follows core defense strategy where cluster-head is the centre point to defend intruder and concentrates on saving the power of sensor nodes by distributing the responsibility of intrusion detection to three layer nodes.
- In [7], a Hybrid Intrusion Detection System (HIDS) has been proposed in heterogeneous cluster based WSN (CWSN).The attacks such as spoofed, altered, or replayed routing information, sinkhole, sybil, wormholes, acknowledgment spoofing, select forward, hello floods can be detected using this model.
- In [8], a hierarchical model (three layer architecture) is proposed based on weighted trust evaluation (WTE) to detect malicious nodes by monitoring its reported data.
- In [9], a dynamic model of intrusion detection (DIDS) has been proposed for WSN. This is a hierarchical model of IDS based on clustered network to battle the low energy. It can use distributed defense which has the advantage of detecting multiple intruders, albeit, with an increased rate of energy consumption with increase in cluster size.

## 3   Comparative Analysis of Recent ID Schemes

Table 1. STRENGTH, WEAKNESS and FUTURE SCOPE of EXISTING IDS

| Existing IDS | Strength | Weakness | Future Scope |
|---|---|---|---|
| Semantic IDS[1] | 1) Agent node stores the whole ontology in its memory. <br> 2) Energy efficient | 1) Mapping of security ontology with sensor data is vague. <br> 2) Decision making function is not clearly specified. | Algorithms can be improved by using more complex semantics of security ontology. |
| Simple Learning Automata based IDS [2] | 1) Distributed nature avoids all other nodes being sacrificed when a single node is affected.<br>2) Energy efficient<br>3) Self-learning nature optimizes packet sampling efficiency. | Computational complexity increases because of using dynamic topology by distributed self-learning automation technique. | S-LAID solution can be tested in different application domains of sensor network. |
| Location Aware Trust based IDS [3] | 1) Reputation-based monitoring facilitates detection and isolation of malicious nodes efficiently.<br>2) Location awareness enhances integrity. | Use of encryption algorithm consumes more energy. | Location verification protocol can be extended. |
| Isolation Table based IDS [4] | Primary experiment proves that ITIDS can prevent attacks effectively in terms of live nodes and transmission accuracy. | When the remaining nodes decrease, the intruders can penetrate WSN more easily. | Anomaly detection technique can be extended for improvement. |
| Ranger based IDS [5] | 1) Intruder can not attack WSN through isolated anomalous nodes.<br>2) Lightweight model works in energy-efficient manner. | It mainly focuses on Sybil attack. | It can be implemented through standard protocols (e.g. Zigbee) for performance evaluation. |
| Hierarchical Overlay Design based IDS [6] | 1) Reliability, efficiency and effectiveness are high for a large geographical area.<br>2) Distributed four level hierarchy results in highly energy saving structure.<br>3) ID becomes very fast and effective. | 1) IDS needs to wait for intruders to reach the core area whereas nodes can be captured at any area without any notice.<br>2) Total cost of network set up is increased for using policy based mechanism. | Election procedure can be implemented; IDS scalability and definition of detection policy need to be determined, more specifically. |
| Hybrid IDS [7] | 1) Its detection rate and accuracy are high for using hybrid approach. Decision making model is very simple and fast.<br>2) Cluster head is used to reduce energy consumption, amount of data in the entire network and to increase network lifetime. | Rules in the anomaly detection model are defined manually, so performance can not be verified through simulation. | Feature selection in anomaly detection can be done by data mining; Rule based approach can be extended to provide anomaly detection model with better performance and flexibility. |
| Weighted Trust Evaluation based IDS [8] | 1) It detects misbehaved nodes accurately with very short delay.<br>2) Light-weight algorithm incurs little overhead. | It gives rise to high misdetection rate. | More detailed analysis regarding the performance will be studied in the ongoing research. |
| Dynamic Model of IDS [9] | 1) It has remarkable improvement in security, stability and robustness as compared to static IDS. Distributed nature of this model increases security and network's lifetime.<br>2) Upgradation of defense structure increases flexibility. | 1) It needs more time to detect all intrusions.<br>2) Distributed detection consumes more energy. | It can be tested with real life applications to ensure perfectness of the model. |

**Table 2. Analysis of Some of the Recent IDS for WSN**

| Intrusion Detection System | Featurewise differences | | |
|---|---|---|---|
| | **Node Density** | **Detection Rate** | **Energy consumption** |
| S-LAIDS [2] | Node density medium. | Penalty threshold of 0.2 detects 63 to 71% malicious packets, that of 0.8 is able to detect 25 to 33% malicious packets. | Both the reward and the penalty functions are calculated on basis of the residual energy. Removal of malicious node requires less energy. |
| Location aware trust based IDS [3] | Number of sensor nodes within 5 to 100 are deployed randomly in 50 $m^2$ area. | Probability of compromised node detection is certain when the number of neighboring nodes is 15 or less. As the number of neighboring nodes increases, the probability of blacklisting decreases. | No evaluation regarding energy consumption is found. |
| ITIDS [4] | 200 sensor nodes are deployed uniformly within 10000 square meters area. | 95% detection accuracy is achieved when number of monitor nodes equals to 100. | Energy consumption is less for WSN having 50 nodes compared to 100 or 200 nodes. |
| HIDS [7] | Node density is not specified. | 99.81% detection rate, 0.57% phantom intrusion rate and 99.75% accuracy are achieved. Individual detection rate is very low when the training sample is not substantial. | Its energy consumption is very low. |
| WTE based IDS [8] | Number of nodes are within a range from 9 to 900. It has high scalability. | Detection is terminated after more than 25% of all nodes are detected as malicious nodes. Weight penalties values in the range of 0.04 -0.1 can improve detection rate with low misdetection rate. | No evaluation regarding energy consumption is found. |
| DIDS [9] | 70 nodes within transmission range of 4 to 15 m, having cluster size equals to 10 for the overall area of 80m *100m. | When number of nodes equals to 20, all types of defenses can detect intrusion, but when number of nodes is greater than or equal to 40, only distributed defense can detect intrusion. DIDS detection rate is higher within smaller range (90% with a range of at least 15m). | If consumed energy in any node is greater than or equal to 30% before activation of IDS, it can not be selected. Distributed defense results in high energy consumption. The lowest energy in DIDS is about 57%, which is 17% higher than that in SIDS. DIDS can prolong the lifetime of network by 8% on average. |

## 4   Proposed Model

Our objective is to detect the sleep deprivation attack in sensor network. In this section, a lightweight model, **INSOMNIA MITIGATING INTRUSION DETECTION SYSTEM (IMIDS)** is proposed for heterogeneous wireless sensor network (HWSNET) to detect insomnia of stationary sensor nodes. It uses cluster based mechanism in an energy efficient manner to build a five layer hierarchical network to enhance network scalability, flexibility and lifetime. The low energy constraints of WSN necessitate the use of a hierarchical model for IDS. We divide sensor network into clusters which are again partitioned into sectors.

It will minimize the energy consumption by avoiding all the nodes needing to send data to a distant sink node. It uses anomaly detection technique in such a way so that phantom intrusion detection can be avoided logically.

### 4.1 Assumptions

- A sensor can be in any one of the following states:

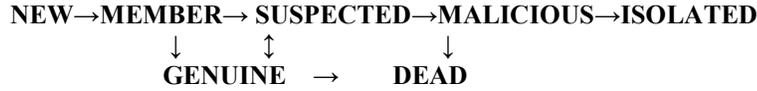

NEW→MEMBER→ SUSPECTED→MALICIOUS→ISOLATED
↓      ↕           ↓
GENUINE  →    DEAD

- Each sensor node has a unique id in the network.
- Each member node has authentic wake-up token.
- A protocol is used to assign a secure wakeup and sleep schedule for the sensor nodes.
- Sink node is honest gateway to another network.
- The threshold values are pre-calculated and set for the entire network.
- If any of cluster coordinator, forwarding sector head, sector monitor or sector coordinator is found to be compromised, reconfiguration procedure takes place dynamically.
- Sensor nodes excluding leaf nodes and forwarding sector heads in the system participate in intrusion detection process.
- Generally, sector coordinator is responsible for anomaly detection and sector monitor is responsible for detection of intrusion.
- Initially, probability of sleeping schedule and wake-up schedule are same ($p$=0.5) for any normal node.
- Initially, trust value of each node is represented by a nibble $t_3 t_2 t_1 t_0$ containing all 1's, belief is set to 1.
- SM may be more than one within a sector.
- SN selects CC and CC selects SC, SM, FSH.
- Anomaly can be detected on the basis of energy consumption rate, allotted wakeup schedule, authentic wakeup token, number of packets received within a time interval. Reputation of sensor node needs to be considered during intrusion detection.

### 4.2 Data Definition

- *Definition 1: Leaf Node LN*– A node N is defined to be a *Leaf Node* if $Child_N\{\ \}=\{\varnothing\}$ AND $Parent_N\{\ \} \neq \{\varnothing\}$. Its detection power(DP) ←0.
- *Definition 2: Setor Coordinator SC* – A node N is defined to be a *Sector Coordinator* if $Rem\_eng_N = MAX\_ENG\{FN[\ ]\}$, where FN[] → follower nodes.
- *Definition 3: Setor Monitor SM* - A node N is defined to be a *Sector Monitor* if $DP_N = MAX\_DETECT\{N[\ ]\}$, where $N \notin \{CC_k, SN\}$ AND $DP_N$ → power required by a node for intrusion detection.
- *Definition 4: Forwarding Sector Head FSH* - A node N is defined to be a *Forwarding Sector Head*, where $hop\_distance_N\{\} = min\{hop\_distance_N\ from\ CC_k\}$, where $N \notin CC_k$. Its detection power (DP) ←0.
- *Definition 5: Cluster Coordinator CC* - A node N is defined to be a Cluster Coordinator, if $Rem\_eng_N = MAX\_ENG\{N[\ ]\}$ AND $CAPACITY_N = MAX(CAPACITY_N)$, where $N \notin SN$ AND $CAPACITY_N = (DEGREE_N/INITIAL\_ENG_N)*Rem\_Eng_N$. $DEGREE_N$→number of nodes within its radio range.
- *Definition 6: Sink Node SN* - A node N is defined to be a *Sink Node* if $Child_N\{\ \} \neq \{\varnothing\}$ AND $Parent_N\{\ \} = \{\varnothing\}$.

### 4.3 System Model

Figure 1 describes the main building block of the system model. Here SN–> SINK NODE; CC–>CLUSTER COORDINATOR; SM–>SECTOR MONITOR; FSH–>FORWARDING SECTOR HEAD; SC–>SECTOR COORDINATOR**;** LN–> LEAF NODE;

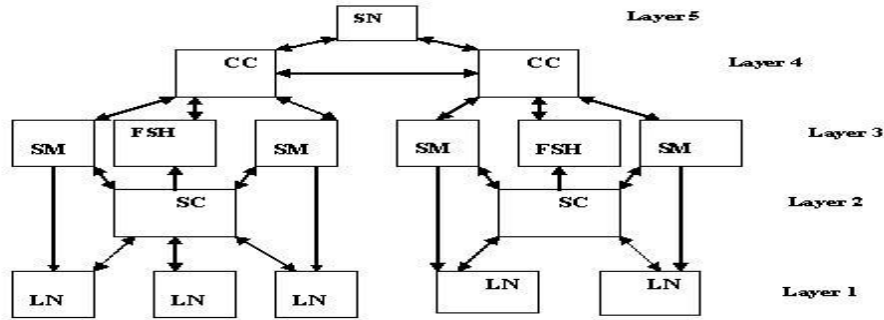

**Fig. 1.** Layered Model

### 4.3.1 Description of Each Layer

The five layers of sensor network are described below-
- **Layer 1**: In this lowest layer leaf nodes sense environmental data and send it to its immediate next higher layer i.e. layer 2. Layer 1 has no anomaly detection capacity.
- **Layer 2**: This layer includes sector coordinator (SC) of each sector that collects data from layer 1 and checks for anomaly. Sector coordinator maintains membership list [] of all leaf nodes within a sector, normal profile [] (tuple space that consists of sensor node's attribute) and knowledge base [] (system parameters, application requirements). Suspected nodes are penalized and legitimate nodes are rewarded by SC. Reputation list [] is updated. Suspected node details are inserted in suspected list [] before forwarding to SM and valid packets are forwarded to FSH of layer 3.
- **Layer 3**: This layer includes sector monitor(SM) and forwarding sector head (FSH). Sector monitor maintains suspected list [], normal profile [], knowledge base [], reputation list []. SM can detect intruders, compromised nodes and isolate them by inserting the details into quarantine list [] and forwards the information to cluster coordinator (CC).FSH (nearest neighbor of cluster coordinator) acts as router that inserts valid packet details to forwarding table [] and forwards valid packet of legitimate nodes to CC of layer 4.
- **Layer 4**: This layer constitutes the cluster coordinator (CC) which controls SM and FSH of each Sector within a cluster. It inserts valid packets details to valid list [] and forwards data to the sink node. Cluster coordinators (CC) can cooperate with each other to form global IDS.CC contains backup copy of its own cluster.
- **Layer 5**: The topmost layer is the sink node that collects data from lower layer and it acts as a gateway between sensor network and other networks or acts as access point. SN contains backup copies of all clusters.

### 4.3.2 IMIDS : Insomnia Mitigating Intrusion Detection System

The entire heterogeneous sensor field is divided into overlapping or disjoint clusters like $C_k$, for $k \in \{1,..,r\}$, r being the number of clusters in the sensor network. Each cluster consists of its member nodes including a cluster coordinator (CC). Let $mem_1, mem_2, ....,mem_n$ be the members of a cluster $C_k$, which are unaware of their locations and n is the number of members within a cluster excluding CC. Clusters are partitioned into non-overlapping sectors like $S_j$, for $j \in \{1,…,m\}$, where m is the number of sectors within a cluster, where $r<<m$. We assume three types of sensor nodes in this five layered model: (i) leader nodes or LDN (in layer 3 and 4) (ii) follower nodes or FN (in layer 1 and 2) and (iii) sink node or SN (in layer 5). Leader nodes can be equipped with EXIDS (extended IDS), but only the node designated as sector monitor can activate it. Cluster coordinator (CC) and sink nodes (SN) are also using EXIDS for detecting intrusion during its requirement. SIDS (simple IDS) can be loaded in all follower nodes, but can be activated only at sector coordinator of layer 2 for detecting anomaly. Sector coordinator collects sensing data within allotted TDMA time slot of each leaf node in a sector. Sector coordinator (SC) monitors the sensor nodes for detecting anomaly by SIDS. Suspected nodes are penalized and legitimate nodes are rewarded. Forwarding sector head (FSH) forwards valid packets to CC. Sector monitor (SM) decides whether a suspected node is malicious or not. EXIDS has the responsibility to declare the suspected node as malicious and to drop fake or corrupted packets. To avoid phantom intrusion detection logically, suspected nodes get chance to increase their reputation by SM, if it is not decided as malicious. Intruder/Malicious nodes are isolated in quarantine list; so that no intrusion occurs through these nodes.

If HWSNET is considered as a graph G (V, E), any edge E between two nodes $n_i$ and $n_j$ is valid if and only if distance between two nodes $D_{i,j} <= R_{tr}$ (transmission range). Detection power of LN and FSH are 0%, SC,CC,SN are 50% and SM is 80%. When detection power reaches to minimum threshold, detection capacity is automatically disabled. Reconfiguration procedure takes place dynamically if any node found to be suspected i.e. energy consumption rate greater than normal consumption rate. Each of leader and follower nodes must be included within a cluster. If any node is under more than one CC then RSSI value need to be checked. If there is a tie, it is broken randomly. Anomaly is detected by SC. But there is a possibility of false positive or false negative. If any genuine node is suspected by SC (false positive), SM can detect it and takes final decision. If any compromised node is treated as genuine and forwarded to FSH (false negative), CC can detect it.

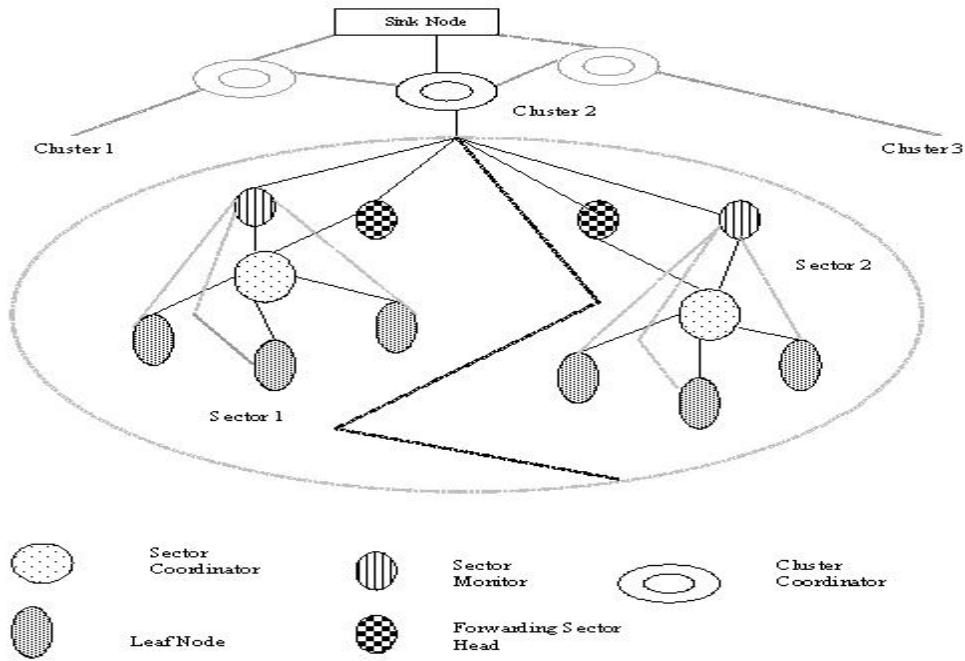

Fig. 2. Cluster Based Heterogeneous Sensor Network

### 4.3.3 Procedural Steps of IMIDS

- **Initialization Phase:**
  Sensor nodes are deployed in the sensor field during this phase. A unique identification number consisting of the geographical position vectors is assigned to each new node. The sink node searches

- **Cluster Coordinator Selection and Cluster Formation Phase:**
  for its neighbors to acquire energy details of all nodes after broadcasting advertised message.
  Cluster coordinator is selected among all leader nodes and its coverage area is considered as cluster. The Cluster-head details are broadcasted to all its neighbors. The neighbor nodes collect advertised messages during a given time interval and send a join message to nearest cluster coordinator for all nodes within the range of any specific cluster coordinator. Intersection of two cluster domains may or may not be NULL.
- **Sector Coordinator Selection and Sector Formation Phase:**
  Sector coordinator is selected among all follower nodes and its detailed information along with node-id is broadcasted to all of its neighbors. Its coverage area is considered as sector. Intersection of domains of two sectors must be NULL. Sector monitors and forwarding sector heads are selected for each sector.
- **IDS Activation Phase:**
  Activate IDS preinstalled in cluster coordinators, sector monitors and sector coordinators.
- **Reconfiguration Phase:**
  When cluster coordinator or sector coordinator's behavior deviates from normal, reconfiguration procedure takes place.
- **Data Transfer and Intrusion Detection Phase:**
  After sector coordinator selection is done each follower node (leaf node) sends data to the sector coordinator that transfers genuine packet to its cluster coordinator through forwarding sector head. Cluster coordinator collects valid data from all sectors within its coverage area and then forwards aggregated packets to sink node.

### 4.3.4 Selection Procedure

(i) Sink node broadcasts its node-id and query message to acquire current residual energy of each sensor node within its coverage area.
(ii) According to response from sensor nodes, sink node categorized sensor nodes into leader nodes having high energy and follower nodes having comparatively low energy.
(iii) Leader node having minimum distance from sink node, maximum residual energy among all other leader nodes and high reputation, is selected as cluster coordinator.
(iv) Remaining leader nodes within a cluster having high detection power is selected as sector monitor, whereas leader nodes having minimum distance from cluster coordinator is selected as forwarding sector head.
(v) Follower nodes having maximum energy among all follower nodes are selected as sector coordinator, other follower nodes are considered as leaf nodes.

## 5 Performance Analysis

In this section, we validate our analysis using simulation in MATLAB. Performance has been studied by simulating sensor nodes in the existing ITIDS and proposed IMIDS. In figure 3, the result of the simulation shows that number of alive nodes with respect to increasing time in second is more in IMIDS. Therefore it can be said that HWSNET lifetime is better by using IMIDS than ITIDS. Because IMIDS uses dynamic configuration and cluster is further partitioned into sectors. In figure 4, the result shows that accuracy is comparatively high in IMIDS because here sector monitors which have high detection power are used to detect intrusion; whereas in ITIDS low energy member nodes are considered as monitor nodes. In figure 5, energy consumption is compared with respect to the density of sensor nodes with clusterization and sectorization and without clusterization or sectorization. The result of simulation shows energy consumption is comparatively less when sensor field is partitioned into clusters and sectors. After analyzing performance, it can be said that proposed IMIDS can prolong network lifetime, detect intrusion accurately and consumes less energy to mitigate sleep deprivation attack.

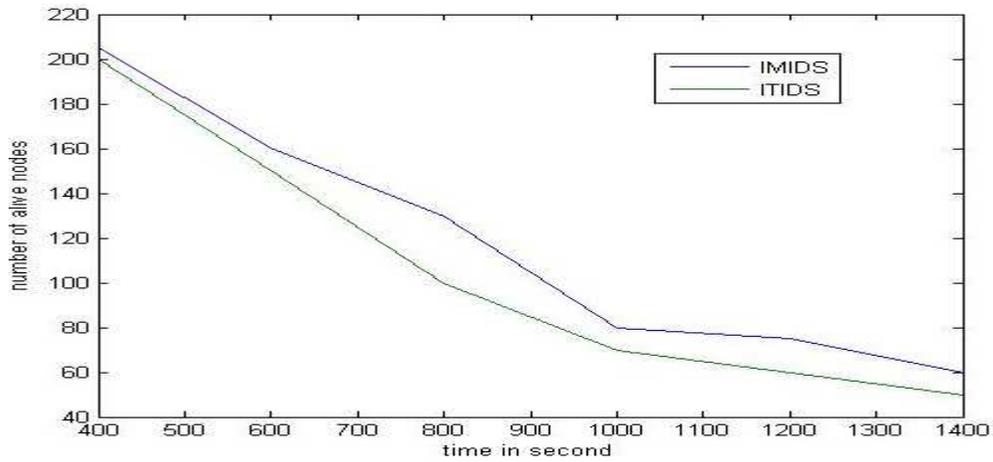

**Fig. 3. Comparison of the number of alive nodes between ITIDS and IMIDS**

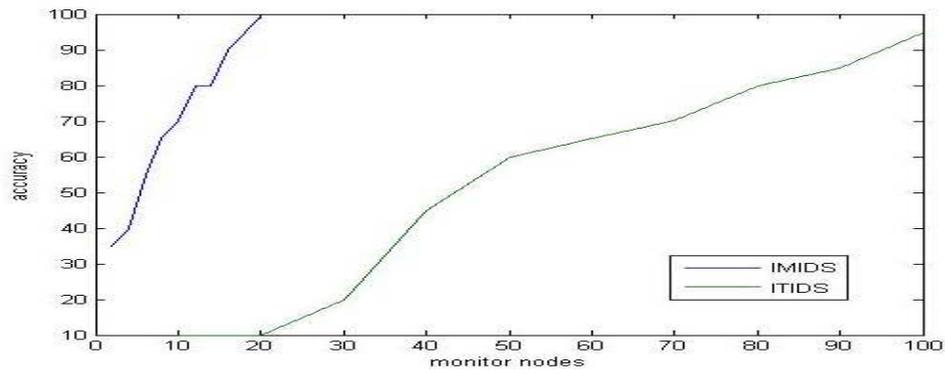

**Fig. 4. Comparison of the accuracy between ITIDS and IMIDS**

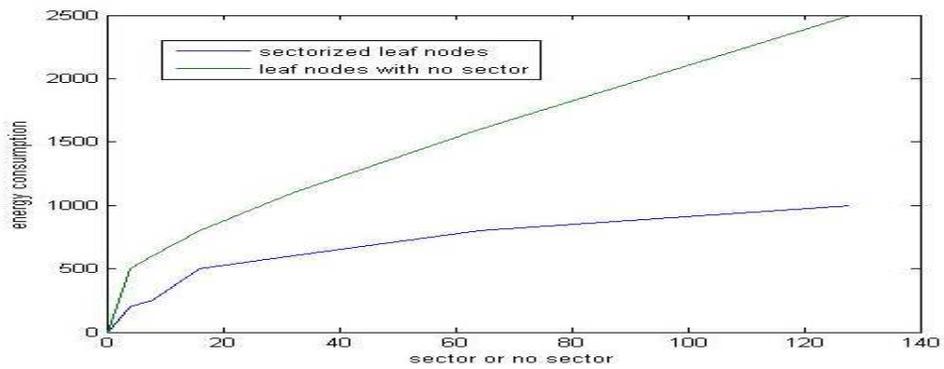

**Fig. 5. Energy consumption with sectorization and without sectorization in IMIDS**

# 6 Conclusion

In this paper, we present a review of recent works on different approaches of IDS for WSN. It has been observed that these intrusion detection systems are not adequate for protecting WSN from intruders efficiently. The need of the day is an IDS for detecting intrusions accurately in an energy-efficient manner. Among the different types of prevalent attacks, sleep deprivation attack at link layer has been found to be the most devastating one for sensor nodes, exhausting the battery life very quickly. This paper comes up with the idea of a novel IDS that can mitigate sleep deprivation attack without using MAC based protocols like S-MAC, T-MAC, B-MAC, G-MAC. The outline of layer based approach using cluster technique to design a lightweight IDS capable of detecting insomnia of sensor nodes with less energy consumption has been documented here. The aim of this proposed model is to extend the lifetime of the WSN, even in the face of sleep deprivation attack. Generally, intruder attacks lower layer leaf nodes in HWSNET. In this model, intrusion detection is mainly focused on layer 1 that has no intrusion detection capacity of its own. Simulation proves the effectiveness of proposed model. At present work is on for more detailed analysis of IMIDS in a simulated environment.